\pdfoutput=1
\documentclass[prb,twocolumn,showpacs,amsmath,amssymb,floatfix,superscriptaddress]{revtex4}

\usepackage{graphicx}
\usepackage{dcolumn}
\usepackage{bm}
\usepackage{amsmath}
\usepackage{amssymb}
\newcommand*{\Tr}{%
	\mathrm{Tr} }

\begin{document}
	
	\title{First Principles Study of Angular Dependence of Spin-Orbit Torque in Pt/Co and Pd/Co Bilayers}
	
	\author{Farzad Mahfouzi}
	\email{Farzad.Mahfouzi@gmail.com}
	\affiliation{Department of Physics and Astronomy, California State University, Northridge, CA, USA}
	\author{Nicholas Kioussis}
	\email{Nick.Kioussis@csun.edu }
	\affiliation{Department of Physics and Astronomy, California State University, Northridge, CA, USA}
	
	\begin{abstract}
	 Spin-orbit torque (SOT) induced by spin Hall and interfacial effects in heavy metal(HM)/ferromagnetic(FM) bilayers has recently been employed to switch the magnetization direction using in-plane current injection. In this paper, using the Keldysh Green's function approach and first principles electronic structure 
	 calculations we determine the Field-Like (FL) and Damping-Like (DL) components of the SOT for the HM/Co (HM = Pt, Pd) bilayers.  Our approach yields the angular dependence of both the FL- and DL-SOT on the magnetization direction without assuming a priori their angular form. Decomposition of the SOT into the Fermi sea and Fermi surface contributions reveals that the SOT is dominated by the latter. Due to the large lattice mismatch between the Co and the HM we have also determined the effect of tensile biaxial strain on both the FL- and DL-SOT components. The calculated dependence 
	 of FL- and DL-SOT on the HM thickness is overall in good agreement with experiment. The dependence of the SOT with the position of the Fermi level suggests that the DL-SOT dominated by the Spin Hall effect of the bulk HM.
\end{abstract}

	\maketitle
	
	\section{Introduction}\label{sec:intro}
	Spin-Orbit Torque (SOT) due to inplane current flow in heavy-metal/ferromagnet (HM/FM) bilayers has attracted
	considerable attention in recent years as a method to switch efficiently the magnetization direction of ultrathin FM films at room temperature.\cite{Manchon2008,Miron2011,Liu2012,Liu2012_1,Cubukcu2014,
	CZhang2015,Miron2010}
	Experimental\cite{Miron2011,Cubukcu2014,Miron2010,Garello2013} and
theoretical\cite{Haney2013,Lee2015,Freimuth2014} studies have established that the SOT can be separated into a Field-Like (FL), $T^{FL} \vec{m}\times\vec{y}$, and Damping-Like (DL), $T^{DL} \vec{m}\times(\vec{m}\times\vec{y})$, components, where $\vec{m}$ is the unit vector pointing along the direction of the magnetization and $\vec{y}$ is an in-plane unit vector normal to the applied electric field. In the linear response regime, the FL-SOT affects both the size and direction of the effective magnetic field exerted on the magnetic moment, while the DL-SOT only reorients the effective field and is responsible for the angular momentum transfer between the flowing electrons and the FM, thus modulating the ferromagnetic resonance linewidth of the FM\cite{WZhang2015,Ando2008,Mondal2017}.
	
Experimentally, different techniques including, the adiabatic (low-frequency) harmonic Hall voltage\cite{Pi2010,Hayashi2014,Ghosh2017,MacNeill2017}, the Spin-Torque Ferromagnetic Resonance(ST-FMR) \cite{Kumar2017,Liu2011,Mellnik2014,Goncalves2013,MacNeill2017} and the magneto-optical Kerr effect (MOKE)\cite{Xin2014,Xin2016}, have been used to quantitatively measure the DL- and FL-SOT components.
Furthermore, the adiabatic harmonic Hall voltage  approach has been recently employed to investigate the magnetization orientation dependence of the SOT \cite{Ghosh2017,Qiu2013,Kim2013}.

The experimental observations invite several important questions pertaining to the microscopic origin of: ({\it i}) The bulk versus interfacial SOT\cite{Mahfouzi_PRB2016,Haney2013,Freimuth2014,Wang2016} ({\it ii})  Higher order angular terms of the FL-SOT \cite{Garello2013,Ghosh2017,Qiu2013,Kim2013} and ({\it iii}) The HM thickness dependence of the SOT \cite{Nguyen2016,Ghosh2017}. To address these questions an accurate description of the electronic structure of the bilayer in terms of the crystal structure and the interfacial hybridization of the electron orbitals is necessary, signifying the importance of the first principle study of the SOT phenomena.\cite{Freimuth2014,Filipe2017}	
	\begin{figure}
		\includegraphics[scale=0.4,angle=0]{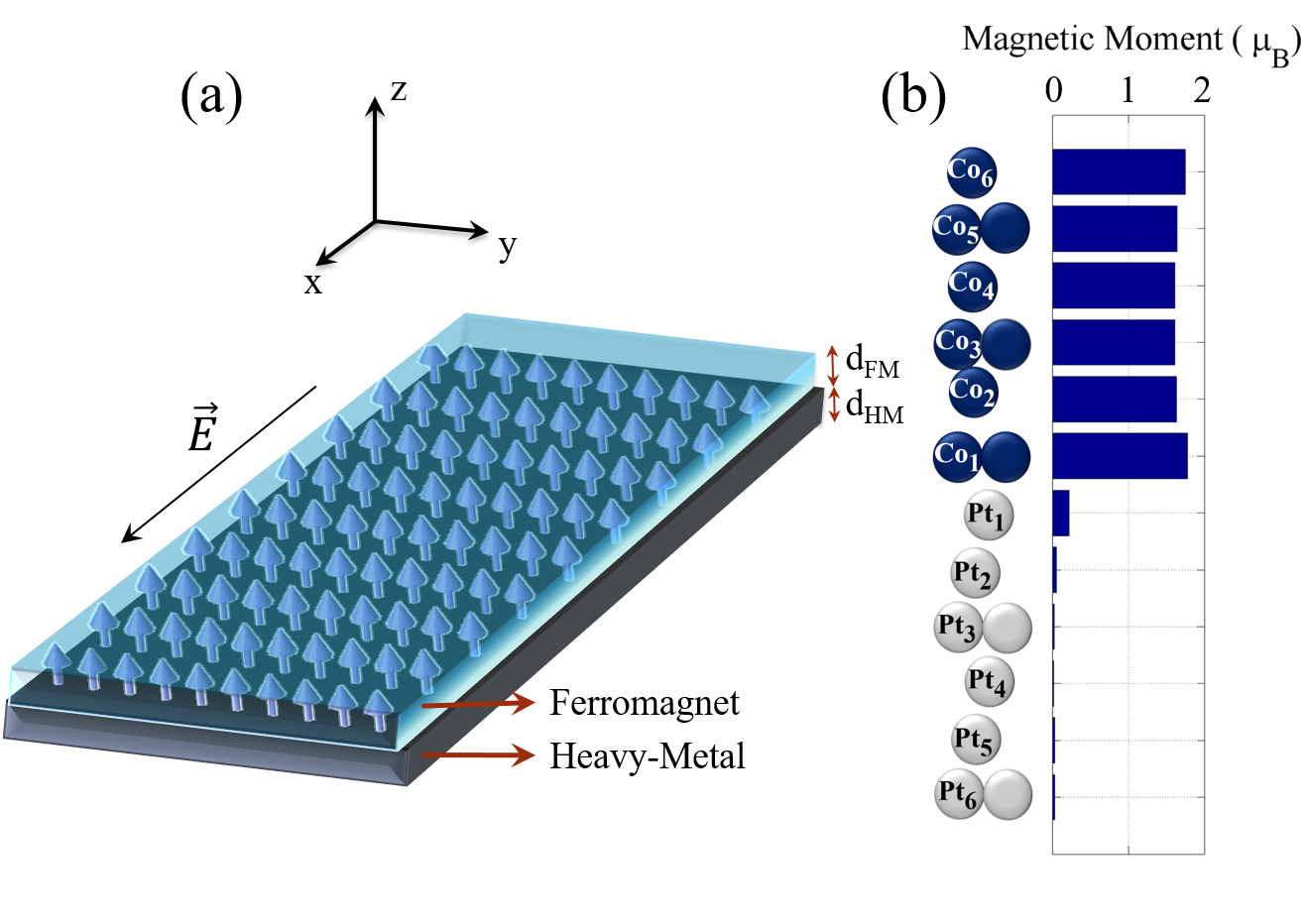}
		\caption{(Color online) (a) Schematic device setup consisting of a Ferromagnet/Heavy-Metal bilayer system under an applied electric field along the $x$-axis. (b) Atomic stacking along the [111] direction of the Co/Pt bilayer slab composed of six layers of hcp Co on six layers of fcc Pt (111). We also show the layer-resolved magnetic moment.}
		\label{fig:fig1}
	\end{figure}

The objective of this work is to employ an {\it ab initio}-based framework which links the Keldysh Green's function approach  with first principles
electronic structure calculations to determine the FL- and DL-SOT of the Co/Pt and Co/Pd (111) bilayers. This approach yields the well-known
angular forms for both SOT components without assuming a priori their angular dependence. We show that the
 DL-SOT can be separated into Fermi sea and Fermi surface contributions where the latter is dominant. We present results of the effect of biaxial strain and of heavy metal thickness on both components of the SOT and compare the {\it ab initio} 
 results with experiment. In agreement with experiment, we find that the FL-SOT extrapolates to a nonzero value with decreasing the HM thickness, indicating its interfacial origin due the Rashba-Edelstein effect (REE). On the other 
 hand, the DL-SOT vanishes
 with decreasing HM thickness suggesting its bulk origin due to the Spin Hall Effect (SHE). This is corroborated by the strong correlation of the
 dependence of the DL-SOT and Spin Hall Conductivity (SHC) on the Fermi level position of the heavy metal.

	\section{Theoretical Formalism}
	The magnetization dynamics of a FM described by a time-dependent unit vector, $\vec{m}$, along the magnetization orientation, is described by the Landau-Lifshitz-Gilbert (LLG) equation of motion,
	\begin{align}\label{eq:LLG}
	\frac{d\vec{m}}{dt}&=-\vec{m}\times\gamma\vec{H}_{ext}+\vec{T}+\alpha\vec{m}\times\frac{d\vec{m}}{dt}.
	\end{align}	
	Here, $\gamma$ is the gyromagnetic ratio, $\alpha$ is the Gilbert damping constant, $\vec{H}_{ext}$  is the external magnetic field and
	 $\vec{T}$, is the current-induced torque on the FM which can be written in the form,
	\begin{align}\label{eq:SOT}
	\vec{T}=\frac{1}{N_kM_s}\vec{m}\times\sum_{\vec{k}}\Tr\Big(\frac{\partial \hat{H}_{\vec{k}}}{\partial \vec{m}}\hat{\rho}_{\vec{k}}\Big),
	\end{align}
	where, $N_k$ is the number of $k$-point mesh for the numerical integration, $M_s$ is the magnetic moment per unit cell in units of Bohr magnetron, $\mu_B$,  $\hat{H}_{\vec{k}}(\vec{m})$ is the electronic Hamiltonian which depends on $\vec{m}$
 and $\hat{\rho}_{\vec{k}}$ is the electronic density matrix.

	In the absence of an external electric field or a time-dependent term in the Hamiltonian, the electronic density matrix is given by $\hat{\rho}^{eq}_{\vec{k}}=\int dEIm(\hat{G}_{\vec{k}E})f(E)/\pi$, where, $\hat{G}_{\vec{k}E}=(E-i\eta-\hat{H}_{\vec{k}})^{-1}$ is the electron Green's function and $f(E)$ is the Fermi distribution function. Here, $\eta=\hbar/2\tau$ is the broadening parameter, where $\tau$ is the relaxation time for the excited electrons. Typically, $\vec{T}^{eq}$ contributes to the exchange interaction between local moments and is responsible for the magnetocrystalline anisotropy arising from the spin-orbit coupling (SOC). To go beyond the equilibrium regime and investigate the effect of the external electric field on the density matrix we employ the Keldysh Green's function formalism\cite{mahfouziThesis2014}, where the density matrix is given by, $\hat{\rho}_{\vec{k}}=\eta\int dE\hat{G}_{\vec{k}E}f(E)\hat{G}_{\vec{k}E}^{\dagger}/\pi$. In the presence of an external electric field, $\vec{E}_{ext}$ the resulting linear drop in the chemical potential, $\delta\mu(\vec{x})=e\vec{E}_{ext}\cdot\vec{x}$, can be taken into account by replacing the energy in the integrand with $E\rightarrow E+i e\vec{E}_{ext}\cdot\vec{\nabla}_{\vec{k}}$, where, we used $\vec{x}=i\vec{\nabla}_{\vec{k}}$ for the position operator. In the linear response regime the nonequilibrium density matrix of the electrons under an external electric field along the $x$ direction is 
	\begin{align}\label{eq:Dens_Mat}
	\frac{\hat{\rho}^{neq}_{\vec{k}}}{eE_{ext}^x}&=\hbar\int \frac{dE}{2\pi i}\left[Im\left(\hat{G}_{\vec{k}E}\frac{\partial\hat{G}_{\vec{k}E}}{\partial k_x}-\frac{\partial\hat{G}_{\vec{k}E}}{\partial k_x}\hat{G}_{\vec{k}E}\right)f(E\nonumber)\right.\\
	&-\eta\left(\left.\hat{G}_{\vec{k}E}\frac{\partial\hat{G}_{\vec{k}E}^{\dagger}}{\partial k_x}-\frac{\partial\hat{G}_{\vec{k}E}}{\partial k_x}\hat{G}_{\vec{k}E}^{\dagger}\right)\frac{\partial f(E)}{\partial E}\right].
	\end{align}

	The first term in the integrand is the Fermi sea contribution originating from the modification of the single electron Green's function due to the electric field and the second term is the Fermi surface contribution. Using the Fermi Surface (FS) contribution of the nonequilibrium density matrix in Eq. (\ref{eq:SOT}) the FS contribution to the current-induced SOT torkance,
	$\vec{\tau}_{FS}=\vec{T}^{neq}_{FS}/eE^x_{ext}$, is given by
\begin{align}\label{eq:FS_DL_SOT}
	\vec{\tau}_{FS}&=\frac{1}{\pi M_sN_k}\vec{m}\times\sum_{\vec{k}}Im\left(\Tr\left[\frac{\partial \hat{H}_{\vec{k}}}{\partial \vec{m}}Im(\hat{G}_{\vec{k}})\hat{v}^{k_x}\hat{G}_{\vec{k}}\right]\right).
	\end{align}

\noindent	Here, $\hat{v}^{k_x}=\frac{\partial \hat{H}_{\vec{k}}}{\partial k_x}$ is the group velocity matrix and the Green's functions are calculated at the Fermi energy, $E=E_F$.
	The torkance can be separated into the field-like ($\vec{\tau}^{FL}$) and a damping-like ($\vec{\tau}^{DL}$) components corresponding to the imaginary and real parts of the Green's function, $\hat{G}_{\vec{k}}=Re(\hat{G}_{\vec{k}})+iIm(\hat{G}_{\vec{k}})$, respectively.
	Note that $\vec{\tau}^{FL}$ ($\vec{\tau}^{DL}$) is even (odd) under $\eta\rightarrow-\eta$ transformation.
	A similar decomposition can also be made from the even and odd components of the torkance under $\vec{m}\rightarrow-\vec{m}$ transformation\cite{Freimuth2014}.
	
	The total non-equilibrium density matrix given by Eq.\eqref{eq:Dens_Mat} can be rewritten in the following form,
	\begin{align}\label{eq:Tot_Dens_Mat}
	\frac{\hat{\rho}^{neq}_{\vec{k}}}{eE_{ext}^x}&=\hbar\int \frac{dE}{\pi}\bigg[Im(\hat{G}_{\vec{k}E})\hat{v}_{k_x}Im(\hat{G}_{\vec{k}E})\frac{\partial f(E)}{\partial E}\nonumber\\
	&-2Im\left(Im(\hat{G}_{\vec{k}E})\hat{v}_{k_x} Re(\hat{G}_{\vec{k}E}^2)\right)f(E)\bigg],
	\end{align}	
	where, the first term leads to the FL-SOT and remains to be a FS quantity only, $\tau_{tot}^{FL}=\tau_{FS}^{FL}$. The second term leads to the total DL-SOT, where, in the ballistic regime, $\eta\rightarrow 0$ and in a representation in which the Hamiltonian is diagonal, we can use $Im(\hat{G}_{\vec{k}E})_{nn}=\pi\delta(E-\varepsilon_{n}(\vec{k}))$ to obtain,

\begin{align}\label{eq:Tot_DL_SOT}
\vec{\tau}^{DL}_{tot}&=\frac{2}{M_sN_k}\vec{m}\times\sum_{nm\vec{k}}Re\left(\frac{Im((\frac{\partial \hat{H}_{\vec{k}}}{\partial \vec{m}})_{nm}\hat{v}^{k_x}_{mn})}{(\varepsilon_{n\vec{k}}-\varepsilon_{m\vec{k}}-i\eta)^2}\right)f(\varepsilon_{n\vec{k}}),
\end{align}

\noindent	where, $n$ and $m$ are band indices and $\varepsilon_{n\vec{k}}$ are the energy bands.
Eq. \eqref{eq:Tot_DL_SOT} is sometimes rewritten in terms of the Berry curvature\cite{Kurebayashi2013,Lee2015,Freimuth2014}. Note, that in contrast to the Fermi surface contribution to the SOT in Eq.\eqref{eq:FS_DL_SOT},  Eq.\eqref{eq:Tot_DL_SOT} yields a nonzero DL-SOT even for vanishing density of states at the Fermi Energy ($Im(G_{\vec{k}E_F})$), as in the case of ferromagnetic/topological insulator heterostructures, where the proximity induced exchange splitting at the interface can open a gap on the surface state of the TI\cite{mahfouzi_PRB2016,mahfouzi_PRB2010,manchonPRB2018}. In Sec. IV we present results for both Fermi sea and Fermi surface contribution to the DL-SOT 
using Eqs.\eqref{eq:FS_DL_SOT} and \eqref{eq:Tot_DL_SOT}, respectively.

	\begin{figure}
		\includegraphics[scale=0.32,angle=0]{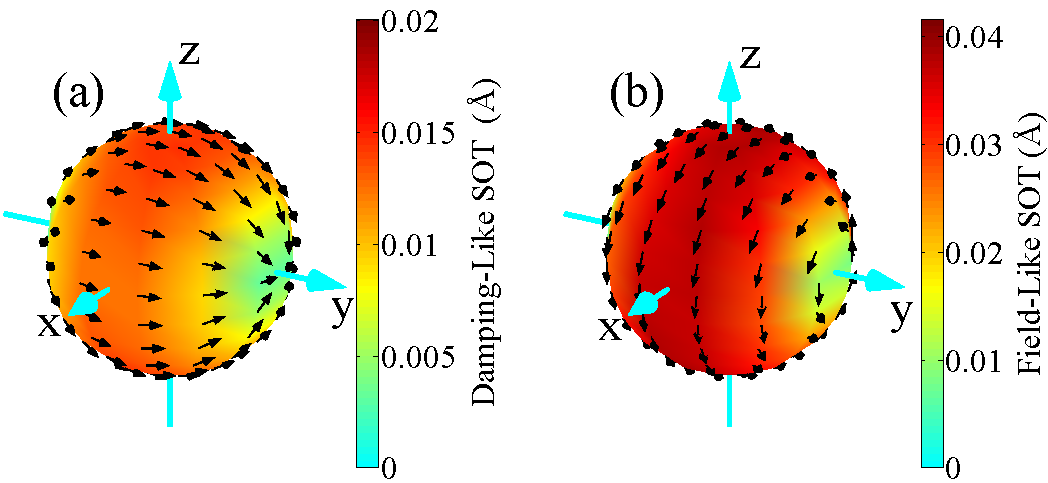}
		\caption{(Color online) Angular dependence of the total (a) DL-SOT,  $\vec{\tau}^{DL}_{tot}$ and (b)  FL-SOT, $\vec{\tau}^{FL}_{tot}$, on
		the magnetization direction, $\vec{m}$, for the Pt(8 ML)/Co(6 ML) 
		bilayer under an external electric field $\vec{E}_{ext}$ along $x$ and for
		broadening parameter, $\eta=28 meV$. The color (arrow) denotes the magnitude (direction) of the SOT for each magnetization direction.}
		\label{fig:fig2}
	\end{figure}
	\section{DFT Calculation}\label{sec:DFT_calc}
	The density functional theory calculations for the hcp Co(0001)/fcc Pt(111) and Co(0001)/fcc Pd(111) bilayers were carried out using the Vienna \emph{ab initio} simulation package (VASP) \cite{Kresse96a,Kresse96b}. The pseudopotential and wave functions are treated within the projector-augmented wave (PAW) method \cite{Blochl94,KressePAW}.
	Structural relaxations were carried using the generalized gradient approximation as parameterized by Perdew {\it{et al.}} \cite{PBE} when the largest atomic force is smaller than 0.01 eV/\AA. As illustrated in Fig. \ref{fig:fig1}(b), the HM($m$)/Co($n$) bilayer is modeled employing the slab supercell approach along the [111] consisting of $m$ fcc Pt or Pd monolayers (MLs) with ABC stacking ($m$=1, 2, $\ldots$, 8) and $n$=6 hcp Co MLs with AB stacking. A 25 \AA~ thick vacuum region is introduced to separate the periodic slabs along the stacking direction ({\it i.e.} $z$-axis). The plane wave cutoff energy is 500 eV and a 14 $\times$ 14 $\times$ 1 $k$ points mesh is used in the 2D Brillouin Zone (BZ) sampling. 
	The in-plane lattice constant of the hexagonal unit cell is set to the experimental value of 2.505~\r{A}  for bulk Co. Furthermore, 
	in order to investigate the effect of epitaxial strain on the SOT we have varied the in-plane lattice constant in the range of 
	$a \in$(2.5, 2.77)~\r{A}, where the latter value corresponds to the bulk Pt lattice constant. 
	Using the tight-binding Hamiltonian obtained from the VASP-Wannier90 calculations \cite{Mostofi} as detailed in Ref.\cite{mahfouziPRB2017_GD}, we have calculated the SOT versus magnetization orientation with a $500\times 500\times 1$ $k$ point mesh for the BZ sampling.
	

	\section{Results and Discussion}\label{sec:Results}
		
	\begin{figure}
		\includegraphics[scale=0.4,angle=0]{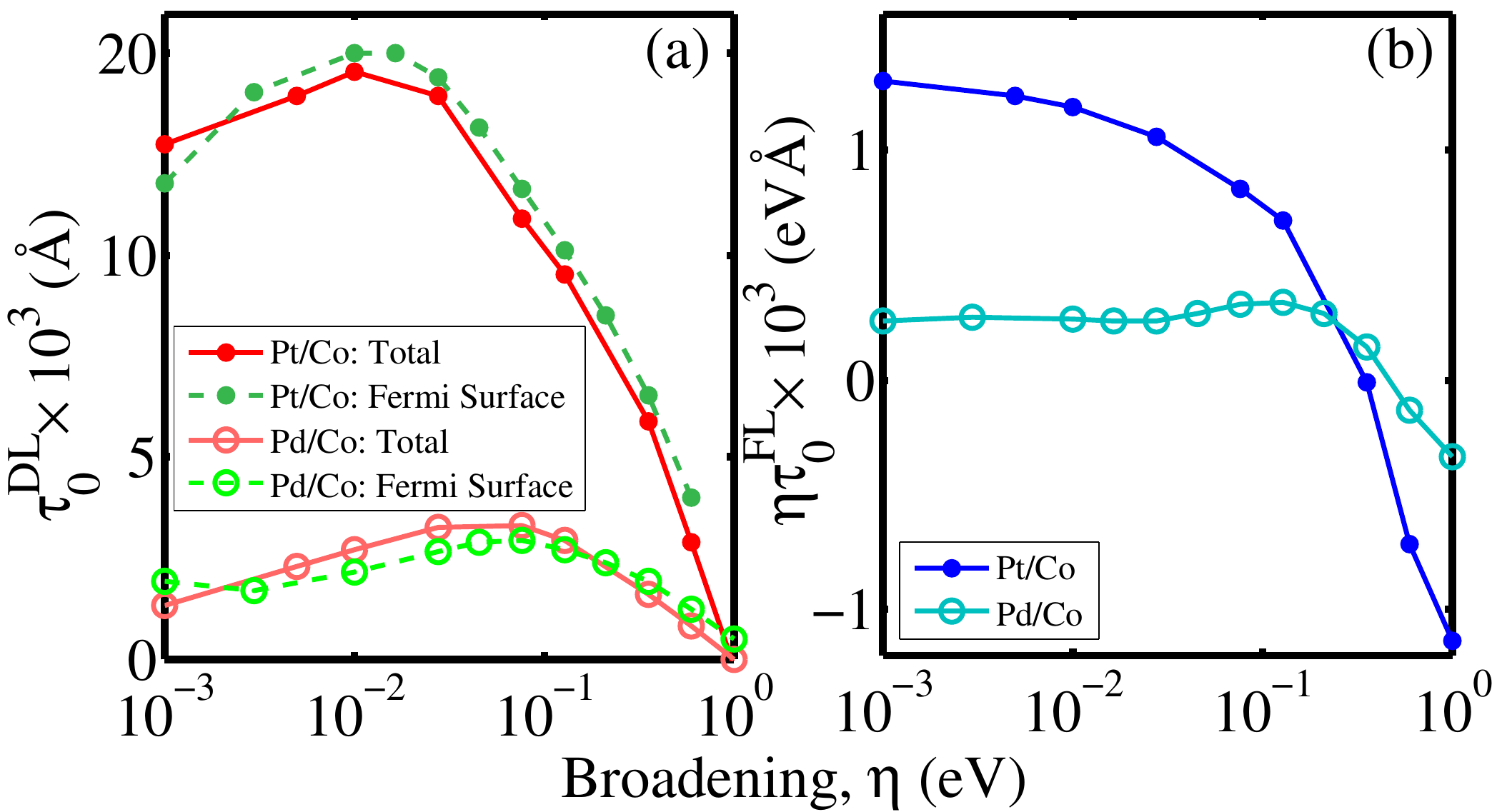}
		\caption{(Color online) (a) Fermi surface contribution (dashed curves) and total (solid curve) DL-SOT and (b) FL-SOT versus broadening parameter, $\eta$ for Pt(8 ML)/Co(6 ML) (filled circles) and Pd (8 ML)/Co (6 ML) (open circles) bilayers, respectively. The total and Fermi surface contribution for the DL-SOT were calculated using Eqs. (\ref{eq:Tot_DL_SOT}) and (\ref{eq:FS_DL_SOT}), respectively.}
		\label{fig:fig3}
	\end{figure}
	%
Figs. ~\ref{fig:fig2}(a) and (b) show the angular dependence of the  $\vec{\tau}^{DL}_{tot}$ and $\vec{\tau}^{FL}_{tot}$ components  for the Pt (8 ML)/Co (6 ML)	calculated from Eqs. (\ref{eq:FS_DL_SOT}) and (\ref{eq:Tot_DL_SOT}), respectively, with $\eta=28 meV$ as an example case that corresponds to a relatively clean system with room temperature broadening
The results show that the angular dependence of the FL-SOT and DL-SOT components follow the expected $\vec{\tau}^{FL}=\tau_0^{FL}(\hat{y}\times\vec{m})$ and $\vec{\tau}^{DL}=\tau_0^{DL}\vec{m}\times(\hat{y}\times\vec{m})$ behavior, without however, a priori assumption of their angular form. Within the accuracy of the calculations, we do not find any contribution from higher-order angular contributions
of the magnetization direction, $\vec{m}$, to the FL-SOT, as suggested by the experimental observations\cite{Garello2013,Ghosh2017}. This is consistent with earlier first principle calculations\cite{Filipe2017}.

	In Figs. \ref{fig:fig3}(a) and (b) we present the FL- and DL-SOT versus the energy broadening parameter, $\eta$, for the Pt (8 ML)/Co(6 ML) and  Pd (8 ML)/Co(6 ML) bilayer systems, using the in-plane lattice constant, {\it a}=2.5\AA~, of bulk Co. The dashed and solid curves in Fig. \ref{fig:fig3}(a) represent the
	Fermi surface [Eq. (\ref{eq:FS_DL_SOT})] and total [Eq. (\ref{eq:Tot_DL_SOT})] DL-SOT, respectively. 
	The broadening parameter,$\eta$, is a phenomenological parameter, which describes the effect of disorder due to impurities, temperature, etc. It depends on the experimental growth conditions and is expected to be different for bulk versus interface local atomic environments.  
	 The results for the SOT amplitude versus $\eta$ presented in Figs. \ref{fig:fig3} show that: ({\it i}) within the numerical accuracy of the calculations, the DL-SOT originates exclusively from electrons on the Fermi surface; 
	 ({\it ii}) $\tau^{DL}_0$ converges to a finite value as $\eta\rightarrow 0$,
	 while $\tau^{FL}_0(\eta)$ diverges as $1/\eta$ demonstrating their intrinsic and extrinsic characteristics, respectively; and ({\it iii}) the DL-SOT exhibits 
	 a non-monotonic dependence with $\eta$ while $\eta\tau^{FL}_0(\eta)$ decreases rather monotonically and changes sign for larger $\eta$ values.

\begin{figure}
				\includegraphics[scale=0.4,angle=0]{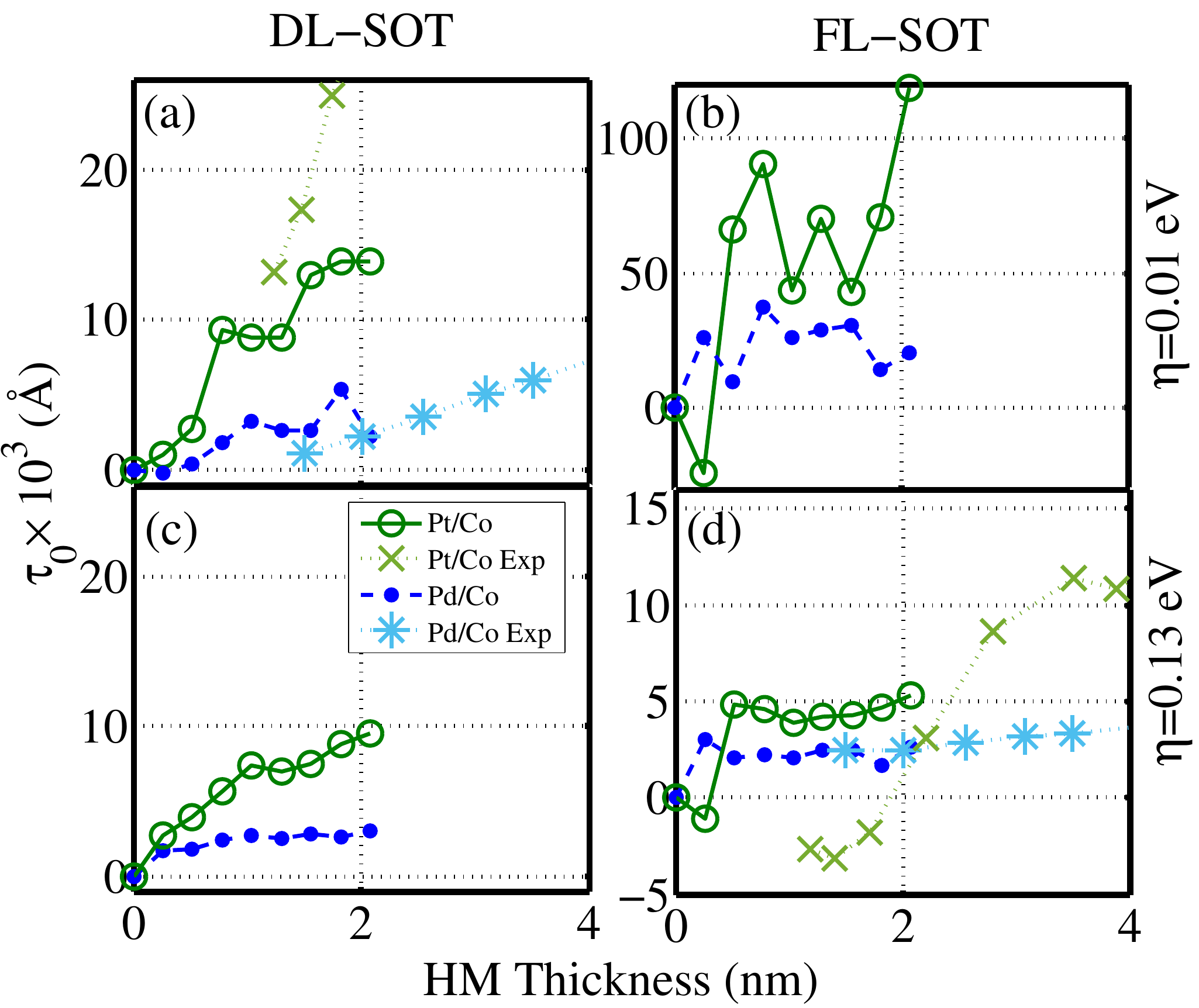}
			\caption{(Color online) Calculated DL-SOT for Pt(n ML)/Co(6 ML) (green circles) and  Pd (n ML)/ Co(6 ML) (blue circles) versus the HM layer thickness for $\eta$ of (a) 0.01 eV (ballistic regime) and (c) 0.13 eV (diffusive regime), respectively. 
			Calculated  FL-SOT for Pt(n ML)/Co(6 ML) (green circles) and  Pd (n ML)/ Co(6 ML) (blue circles) versus the HM layer thickness for $\eta$ of (b) 0.01 eV and (d) 0.13 eV, respectively. 
			For comparison we also show the experimental DL- and FL-SOT
			 results for the Pt/Co\cite{Nguyen2016}  and Pd/Co\cite{Ghosh2017} bilayers.}
			\label{fig:fig5}
		\end{figure}

\begin{figure}
		\includegraphics[scale=0.4,angle=0]{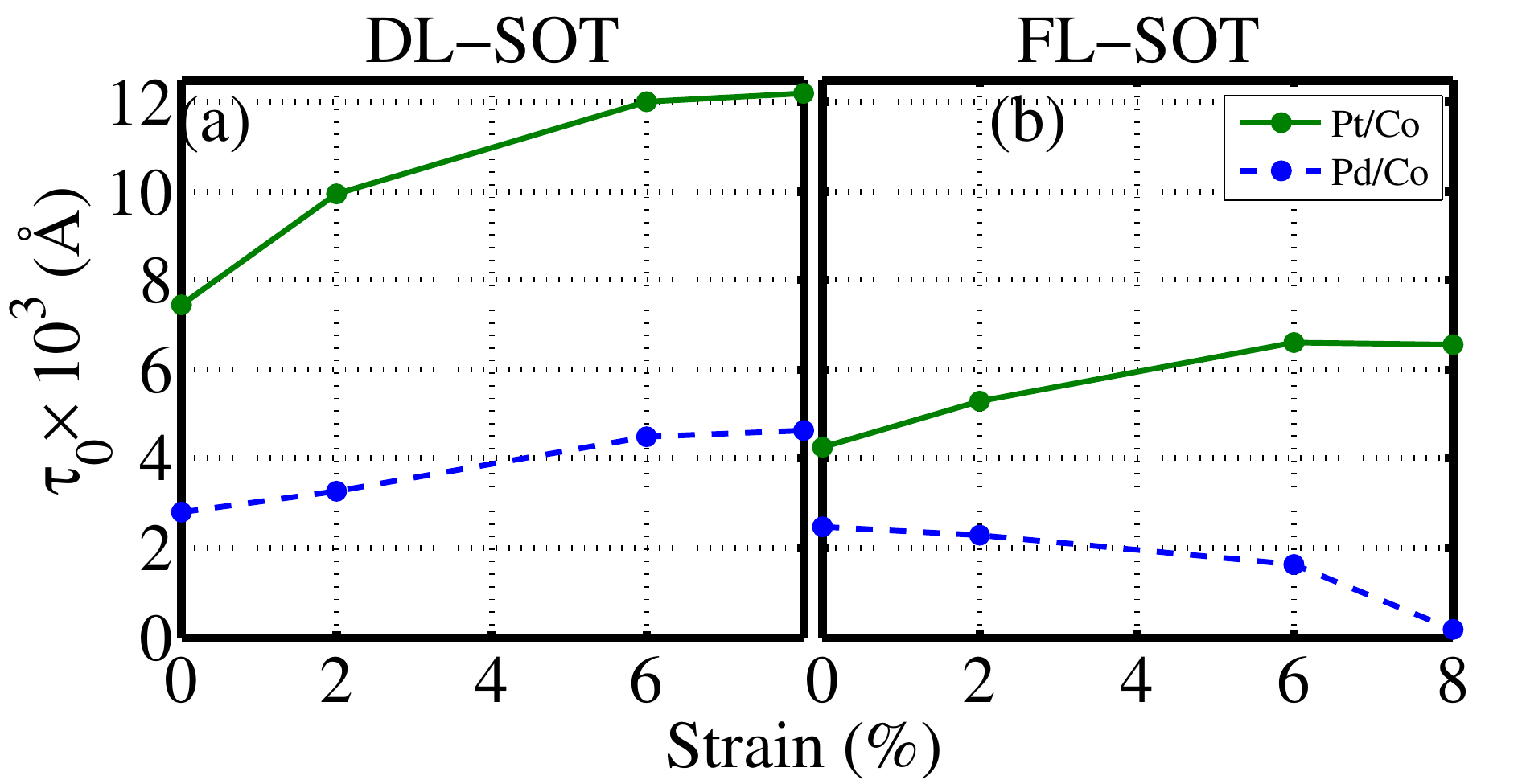}
		\caption{(Color online) (a) FL-SOT and (b) DL-SOT versus tensile biaxial in-plane strain on the Co, for Pt (6 ML)/ Co(6 ML) (solid curves) and  Pd (6 ML)/ Co(6 ML) (dashed curves), respectively, for $\eta=0.13$ eV.}
		\label{fig:fig4}
	\end{figure}

	%
	In order  to elucidate the effect of the HM (Pt, Pd) thickness on the SOT, in  Fig. \ref{fig:fig5} we display the thickness dependence of the DL- and FL-SOT for $\eta=0.01$ eV  [(a) and (b)] corresponding to a relatively clean system and $\eta=0.13$ eV [(c) and (d)] for the \textquotedblleft diffusive\textquotedblright  case within the relaxation time approximation regime. 
	The DL-SOT for the clean system [Fig. \ref{fig:fig5}(a)] exhibits a nonlinear dependence on HM thickness of the form, $\propto1-sech(d_{HM}/\lambda_{HM})\approx \frac{1}{2}\big(\frac{d_{HM}}{\lambda_{HM}}\big)^2,$ for small HM thickness\cite{Ghosh2017}. The agreement with the experimental findings for 
	Pd/Co\cite{Ghosh2017} and Pt/Co\cite{Nguyen2016} is overall good. The {\it ab initio} calculations underestimate the DL-SOT for Pt/Co which may be due to the strain effect shown in Fig. \ref{fig:fig4}(a). On the other hand,  our DL-SOT results for the Pd/Co bilayer yield a smaller spin diffusion length which can be attributed to the violation of conservation laws within 
	the relaxation time approximation. Consequently, the treatment of systems with relatively weak SOC 
	requires a more accurate mechanism\cite{mahfouziPRB2017_GD,Liu2015} of the spin relaxation. 
	For the Pd/Co bilayer both the increase of the FL-SOT with increasing Pd thickness and its saturation as $d_{Pd}\rightarrow 0$ are consistent with experiment indicating its interfacial origin\cite{Ghosh2017}. For the Pt/Co bilayer the FL-SOT reverses sign at 1 ML Pt thickness while experiments report a reversal at about 8 ML ($\approx$ 2 nm)\cite{Nguyen2016}.
	
	Since the lattice constant of bulk Pt ($a_{Pt}=2.8 \AA$) and Pd ($a_{Pt}=2.75 \AA$) are larger than that of bulk Co ($a_{Co}=2.5 \AA$), in Figs. \ref{fig:fig4}(a) and (b) we show the strain dependence of the FL- and DL-SOT, respectively, for the Pt(6 ML)/Co(6 ML) and  Pd(6 ML)/Co(6 ML), 
	where the strain, $\epsilon = (a -a_{Co})/a_{Co}$, and the broadening parameter,  $\eta=0.13$ eV,
	corresponding to the diffusive regime which is more pertinent to experiments on ultrathin bilayers. The increase of the DL-SOT in Pt/Co (Pd/Co) bilayer with strain, which is relatively independent of the chosen value for $\eta$, can be attributed to the increase of the SHC  as discussed in the following.
	
\begin{figure}
		\includegraphics[scale=0.4,angle=0]{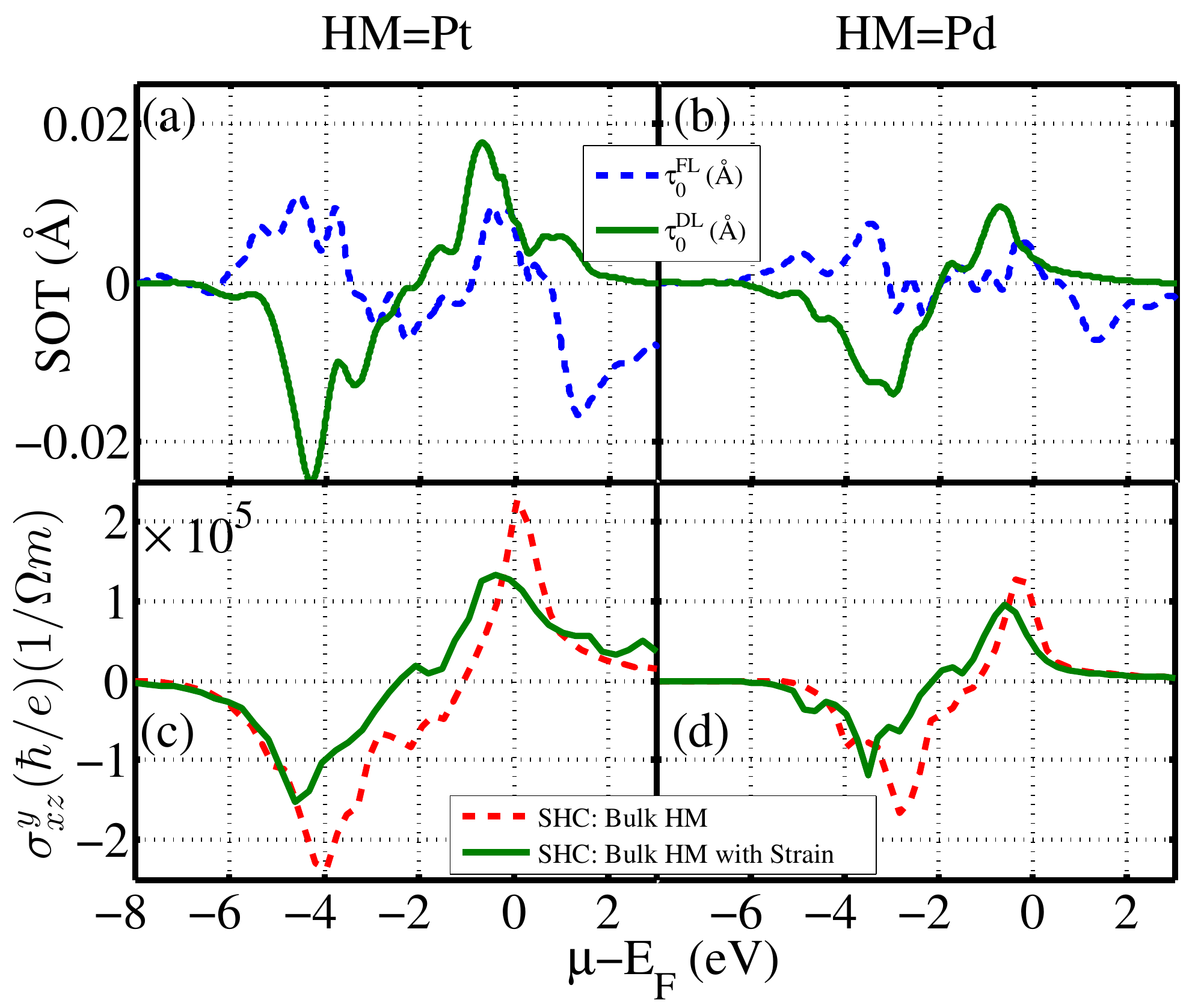}
		\caption{(Color online) FL- and DL-SOT as a function of Fermi level position for (a) Pt (6 ML)/Co (6 ML) and (b) Pd(6 ML)/Co (6 ML).  Spin Hall conductivity, $\sigma_{xz}^y$, as a function of Fermi level position for bulk (c) Pt and (d) Pd. We considered both zero strain as well as -11$\%$(-9$\%$) compressive biaxial strain which lead to 17$\%$(12$\%$) increase of the out-of-plane interlayer distance for bulk Pt (Pd).}
		\label{fig:fig6}
	\end{figure}

	In order to gain an insight into the origin of the SOT and how it changes with shift of the chemical potential,  {\it e.g.} due to doping, in Fig. \ref{fig:fig6}(a) and (b) we present both the FL- and DL-SOT 	for the Pt(6 ML)/Co(6 ML) and Pd(6 ML)/Co(6 ML) bilayers, respectively, as a function of the Fermi level position, $\mu$ -E$_F$ (E$_F$ is the Fermi level) for $\eta=0.1$ eV.
	The FL-SOT in both systems reverses sign as the chemical potential is raised to about +0.1 eV which may be the origin of the experimentally reported sign reversal of the FL-SOT in ultrathin Pt due to electron doping\cite{Nguyen2016}. To understand the origin of the dependence of the DL-SOT on the Fermi level position in  Figs.\ref{fig:fig6} (c) and (d)  we present the SHC of bulk Pt and Pd, respectively,versus the shift of the chemical potential. The SHC is calculated by  replacing $\frac{1}{M_s}\vec{m}\times\frac{\partial\hat{H}_{\vec{k}}}{\partial\vec{m}}$ in Eq. \eqref{eq:Tot_DL_SOT} with $\frac{e^2}{\hbar^2 V_{HM}}I_z^{S_y}$,\cite{Guo2008}
	
	\begin{align}\label{eq:Tot_SHC}
	{\sigma}^{y}_{xz}&=\frac{2e^2}{\hbar^2 V_{HM}N_k}\sum_{nm\vec{k}}Re\left(\frac{Im((I^{S_y}_z)_{nm}\hat{v}^{k_x}_{mn})}{(\varepsilon_{n\vec{k}}-\varepsilon_{m\vec{k}}-i\eta)^2}\right)f(\varepsilon_{n\vec{k}}),
	\end{align}
	where, $I^{S_y}_z=\frac{\hbar}{4}\{\hat{\sigma}^y,\frac{\partial\hat{H}_{\vec{k}}}{\partial k_z}\}$ is the spin current operator and $V_{HM}$ is the volume per unit cell of the bulk HM.
	
	Due to the large in-plane lattice constant mismatch between the Pt (Pd) and Co of about 
	-11$\%$(-9$\%$) we also show in Figs.\ref{fig:fig6} (c) and (d) the SHC of the tetragonally strained bulk Pt and Pd. 	
	Our results for the bulk SHC of $2.1 \times 10^5 (\hbar/e)(1/\Omega m)$ for Pt under zero strain in Fig. 6 is in agreement with the literature values of $1.6 \times 10^5$,  $2.2\times 10^5$, and $2.1 \times 10^5$ in Refs.\cite{Wang2016}, \cite{WZhang2015_2} and \cite{Guo2008}, respectively. Similarly, the bulk SHC of $1 \times 10^5(\hbar/e)(1/\Omega m)$, for Pd under zero strain is in good agreement with the value of $1.05\times 10^5$ in Ref.\cite{WZhang2015_2}.  
	
	We find that the biaxial strain results in a reduction of the maximum SHC values and a shift of the SHC peaks to lower energies. The reduction of the maximum value of the SHC with compressive biaxial strain can be understood as an interplay between the enhancement of the in-plane group velocity, $\hat{v}_{mn}^{k_x}$, and the concomitant larger reduction of the out-of-plane spin current, $I_z^{S_y}$, in Eq. \eqref{eq:Tot_SHC}, due to the increase of the in-plane and decrease of the out-of-plane hopping matrix elements, respectively. This result is also consistent with the effect of strain on the DL-SOT presented in Fig. \ref{fig:fig4}(a).  Figs.\ref{fig:fig6} (c) and (d) also show that the FL-SOT increases for the Pt/Co bilayer while it decreases for the Pd/Co bilayer, suggesting the absence of a universal strain dependence of the FL-SOT with biaxial tensile strain. One can clearly see 
	the correlation of the DL-SOT with the SHC over a wide range of chemical potential shift demonstrating  that the DL-SOT is dominated by the SHE of the bulk HM. 

	\section{Concluding remarks}\label{sec:conclusions}
We have employed  an {\it ab initio}-based framework which links the Keldysh Green's function approach  with first principles electronic structure calculations to determine the FL- and DL-SOT of the Co/Pt and Co/Pd (111) bilayers. 
Without assuming a priori the angular form of the SOT components, we find that the 
dependence of the DL-SOT on magnetization direction is of the form 
$\vec{\tau}_{DL}=\tau_0^{DL}\vec{m}\times(\vec{m}\times \vec{y})$ in agreement with 
experiment, while that of the FL-SOT is of the form, $\vec{\tau}_{FL}=\tau^{FL}_0\vec{m}\times\vec{y}$, which, in contrast to experiment, 
it does not exhibit higher-order angular terms. We show that both the FL- and DL-SOTs are dominated by electrons on the Fermi surface. In Pt/Co both components of SOT increase with increasing tensile biaxial strain on Co, while in Pd/Co only the DL-SOT increases under tensile strain. The DL-SOT decreases quadratically with the HM thickness while the FL-SOT saturates to a finite value in Pt/Co while it reveres sign in Pd/Co.
The dependence of the SOT with the position of the Fermi level 
suggests that the DL-SOT results from the Spin Hall effect of 
the bulk HM.

	\begin{acknowledgments}
		The work is supported by NSF ERC-Translational Applications of Nanoscale Multiferroic Systems (TANMS)- Grant No. 1160504 and by NSF-Partnership in Research and Education in Materials (PREM) Grant No. DMR-1205734.
	\end{acknowledgments}

	
	
\end{document}